\documentclass[12pt]{article}

\usepackage{amsmath,amsthm,amsfonts,graphicx,epsfig,multicol,pdfpages,authblk,times}
\usepackage[format=hang,singlelinecheck=0,font={sf,small},labelfont=bf]{subfig}
\usepackage{pifont}
\usepackage{lineno}
\usepackage{color}

\begin{document}

\title{Multi-Band Petahertz Currents Resolved via High Harmonic Generation Spectroscopy}
\author[1]{Ayelet Julie Uzan\footnote{These authors contributed equally to this work}}
\author[1]{Gal Orenstein$^*$}
\author[2]{\'Alvaro Jim\'enez-Gal\'an}
\author[3]{Chris McDonald}
\author[4,2]{Rui E.F Silva}
\author[1]{Barry D. Bruner}
\author[5, 6]{Nikolai D. Klimkin}
\author[7]{Valerie Blanchet}
\author[1]{Talya Arusi-Parpar}
\author[1]{Michael Kr\"uger}
\author[5, 6]{Alexey N. Rubtsov}
\author[2,8]{Olga Smirnova}
\author[2,9,10]{Misha Ivanov}
\author[11]{Binghai Yan}
\author[3]{Thomas Brabec}
\author[1]{Nirit Dudovich}
\affil[1]{\footnotesize Department of Complex Systems, Weizmann Institute of Science, 76100, Rehovot, Israel}
\affil[2]{\footnotesize Max-Born-Institut, Max-Born Strasse 2A, D-12489 Berlin, Germany}
\affil[3]{\footnotesize Department of Physics, University of Ottawa, Ottawa, Ontario K1N 6N5, Canada}
\affil[4]{\footnotesize Department of Theoretical Condensed Matter Physics, Universidad Autónoma de Madrid, E-28049 Madrid, Spain}
\affil[5]{\footnotesize Russian Quantum Center, Skolkovo 143025, Russia}
\affil[6]{\footnotesize Department of Physics, Moscow State University, 119991 Moscow, Russia}
\affil[7]{\footnotesize Universit\'e  de  Bordeaux -CNRS,  CELIA, UMR5107,  F33405  Talence,  France}
\affil[8]{\footnotesize Technische Universit\"at Berlin, Ernst-Ruska-Geba\"ude, Hardenbergstr. 36A, D-10623 Berlin, Germany}
\affil[9]{\footnotesize Blackett Laboratory, Imperial College London, South Kensington Campus, SW7 2AZ London, United Kingdom}
\affil[10]{\footnotesize Department of Physics, Humboldt University, Newtonstrasse 15, 12489 Berlin, Germany}
\affil[11]{\footnotesize Department of Condensed Matter Physics, Weizmann Institute of Science, 76100, Rehovot, Israel}

\renewcommand\Authands{ and }

\maketitle 

\textbf{Strong field driven electric currents in condensed matter systems open new frontiers in petahertz electronics.  In this regime new challenges arise as the role of the band structure and the quantum nature of electron-hole dynamics have yet to be resolved. Here we reveal the underlying attosecond dynamics that dictates the temporal evolution of carriers in multi-band solid state systems, via high harmonic generation (HHG) spectroscopy. We demonstrate that when the electron-hole relative velocity approaches zero, enhanced quantum interference leads to the appearance of spectral caustics in the HHG spectrum. Introducing the role of the dynamical joint density of states (JDOS) we identify its direct mapping into the spectrum, exhibiting singularities at the spectral caustics. By probing these singularities, we visualize the structure of multiple unpopulated high conduction bands. Our results open a new path in the control and study of attosecond quasi-particle interactions within the field dressed band structure of crystals.}

Induced by the strong field interaction, HHG provides a unique spectroscopic scheme to visualize the coherent evolution of petahertz currents inside solids. Since the first observation \cite{ghimire2011observation}, solid HHG opened a door into the study of the electronic structure and dynamics in crystals \cite{vampa2015all,luu2015extreme,garg2016multi,liu2017high,yoshikawa2017high,hohenleutner2015real}, multiple band dynamics \cite{ndabashimiye2016solid,you2017high,you2017laser,hawkins2015effect} and complex many-body phenomena \cite{silva2018high} in crystalline and amorphous systems \cite{you2017high}. For a moderate field strength the electron-hole dynamics are often described semi-classicaly by a single valence and conduction band of the crystal. As we approach the strong field regime, new fundamental questions arise. What is the role of the band structure in such intense, ultrafast processes? How will electrons and holes interact on extremely short time scales, when they are still mutually quantum coherent? These questions pose some of the primary challenges in the emerging field of strong field interactions in solids.

In this paper we identify the quantum nature of the electron-hole wave-packet in solids, probing its strong-field attosecond dynamics over multiple bands. We observe enhanced quantum interference in the vicinity of Van Hove singularities \cite{ashcroft1976solid,van1953occurrence} and resolve the direct link between the dynamical JDOS and the HHG spectrum. Our study applies HHG spectroscopy in MgO, induced by a $\lambda=1.3 \mu m$ driving laser field. Adding a weak perturbative second harmonic (SH) field modulates the internal dynamics, in a close analogy to a lock-in measurement. This scheme allows us to isolate extremely weak signals and identify the contribution of multiple band excitations, covering a spectral range of up to 30 eV. Our results show unequivocally that the internal dynamics related to HHG are dominated by the interband emission \cite{ghimire2011observation,luu2015extreme,garg2016multi,hohenleutner2015real,vampa2015linking,schubert2014sub,kemper2013theoretical,higuchi2014strong,vampa2014theoretical,golde2008high,tancogne2017impact}, which remains the dominant mechanism even when higher conduction bands are involved. We identify the mapping between the dynamical JDOS at unique regions in the Brillouin zone (BZ) and the HHG spectrum. At Van Hove singularities, spectral caustics are induced, leading to a strong enhancement of the HHG signal. At these critical points, the semi-classical picture fails \cite{vampa2015semiclassical}, imprinting the dynamical quantum nature of the strong field interaction on the HHG spectrum.

The interband HHG mechanism can be viewed as a generalized electron-hole recollision process \cite{vampa2015linking,vampa2014theoretical}, described by a semi-classical analysis \cite{vampa2015semiclassical}. Around the peak of the laser field, an electron tunnels from the valence to the conduction band, forming an electron-hole pair. The laser field subsequently accelerates the pair, leading to their recombination and the emission of XUV radiation. This recollision model maps the semi-classical electron-hole trajectories into harmonic energies. The semi-classical analysis is not strictly limited to one conduction band -- the strong laser field may excite the electron into higher bands, leading to the generation of complex spectral features in the HHG spectrum \cite{ndabashimiye2016solid,you2017high,you2017laser}.

\begin{figure}[h]
\centering
\includegraphics[trim= 200 145 100 170,clip,width=0.85\linewidth]{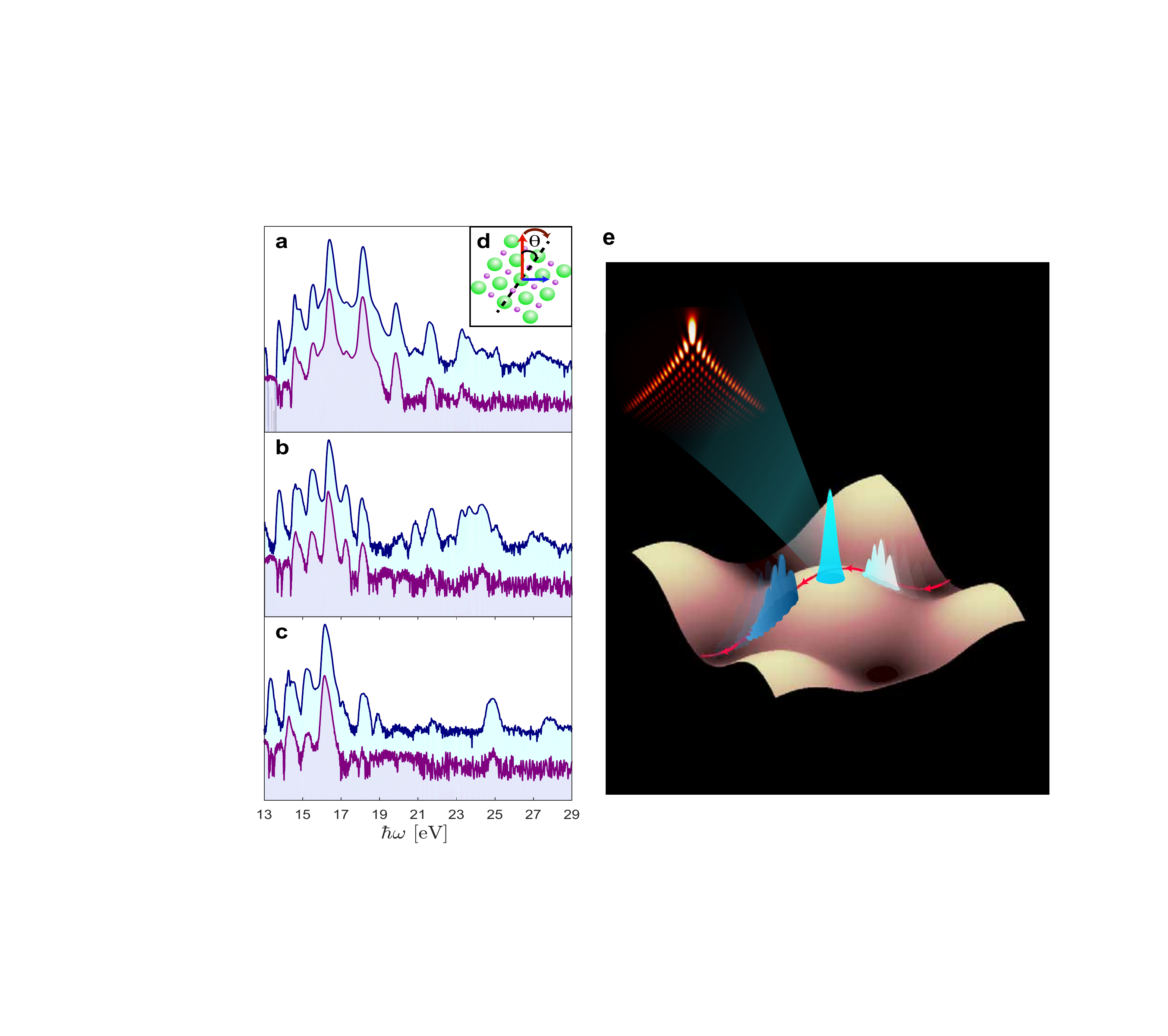}
\caption{\textbf{HHG spectroscopy in MgO}. \textbf{a}, \textbf{b}, \textbf{c} normalized oscillating HHG spectrum (cyan) and the normalized averaged spectrum (violet) in logarithmic scale for orientations \textbf{a} $0^\circ$, \textbf{b} $33^\circ$ and \textbf{c} $45^\circ$. The two spectra are vertically shifted to allow better visibility. \textbf{d} the crystal is oriented to angle $\theta$ with respect to the fundamental field's polarization (red arrow) while the second harmonic (blue arrow) is orthogonally polarized. \textbf{e} Schematic description of enhanced quantum interference, leading to spectral caustics. The white, cyan and blue wave-packets are snapshots of the combined electron-hole wave-function at three different times, probing the band gap of band 10 (figure \ref{fig:fig2}\textbf{d}) along the laser polarization (red line). The modulated white and blue snapshots capture the wave-packet where the electron-hole relative velocity is non-zero. At the extremum of the gap, where the relative velocity is zero (cyan snapshot), enhanced interference leads to the emission of a bright spectral caustic (illustrated by the red interference pattern).}
\label{fig:fig1}
\end{figure}

A fundamental aspect of strong-field induced tunneling is the localization of the excited electron-hole wave-packet in the BZ around the minimum band gap. As the wave-packet evolves under the influence of the laser field, it dynamically probes a narrow stripe of the BZ along the field's polarization, as illustrated in figure \ref{fig:fig1}\textbf{e}. Such localization is a key property of HHG spectroscopy -- enabling the visualization of the band structure with unique angular resolution.

Resolving the angular dependence of the HHG spectrum is extremely challenging -- the HHG signal drops rapidly when the polarization is rotated off the main axes of the MgO crystal \cite{you2017anisotropic}. Such suppression becomes even more significant when the harmonics are produced from higher conduction bands \cite{wu2017orientation}. In order to fully reveal the electron-hole dynamics, both its angular dependence as well as the contribution of multiple bands, we introduce an advanced measurement scheme that enables us to resolve and isolate the weak HHG signals.

Enhancing the detectability of weak signals by subjecting them to a known temporal modulation is a common practice in a wide range of applications, also known as lock-in measurement. We induce such a modulation by adding a weak SH field polarized perpendicular to the fundamental field's polarization. So far this scheme has been applied to probe the internal dynamics in gas phase HHG \cite{bruner2015multidimensional} (and references within) and interband currents in ZnO \cite{vampa2015linking}. Scanning the two-color delay leads to a periodic modulation of the HHG spectrum with four times the fundamental laser's frequency. Fourier analysis isolates the oscillating component, resolving the extremely weak HHG signals, buried in a large experimental background. In addition, XUV emission from higher conduction bands can be highly susceptible to the two-color delay \cite{li2017enhancement}, resulting in an enhanced signature in the oscillating component. Figures \ref{fig:fig1}\textbf{a}, \ref{fig:fig1}\textbf{b} and \ref{fig:fig1}\textbf{c} compare the Fourier amplitudes of the modulated HHG signals and the spectra averaged over the two color delay for crystal orientations of $0^\circ$, $33^\circ$ and $45^\circ$, respectively (see figure \ref{fig:fig1}\textbf{d}). Clearly, the averaged spectrum is accompanied by large background noise and is dominated by harmonic 17 ($\sim$16.2 eV) for $33^\circ$ and $45^\circ$ and harmonics 17 and 19 ($\sim$18.1 eV) for $0^\circ$. Resolving the oscillating spectrum by Fourier analysis shows a dramatic enhancement of the signal over the noise. This measurement exhibits the appearance of new spectral components, which were so far hidden in the averaged spectrum.

The oscillating spectra reveal that for all crystal angles, the spectrum extends beyond the first band edge ($\sim$18 eV), indicating the strong contribution of higher conduction bands. The most important observation is associated with the angular dependence of the measurement -- the oscillating spectrum varies drastically and surprisingly for different angles exhibiting enhanced features and structures over a wide spectral range. We identify three distinct structures: enhanced harmonics around 16-17 eV extending to 18 eV at $0^\circ$, an increased signal in the 20.5-22.5 eV region at $33^\circ$ and a clear spectral feature at 23.5-26 eV, narrowing at $45^\circ$. These features can not be associated with the spectral response of the dipole coupling between the bands, due to their moderate variation with energy (see SI). What is the origin of the enhancement mechanism in these spectral regions and what information does it provide?

\begin{figure}[h]
\centering
\includegraphics[trim=30 135 45 120,clip,width=0.9\linewidth]{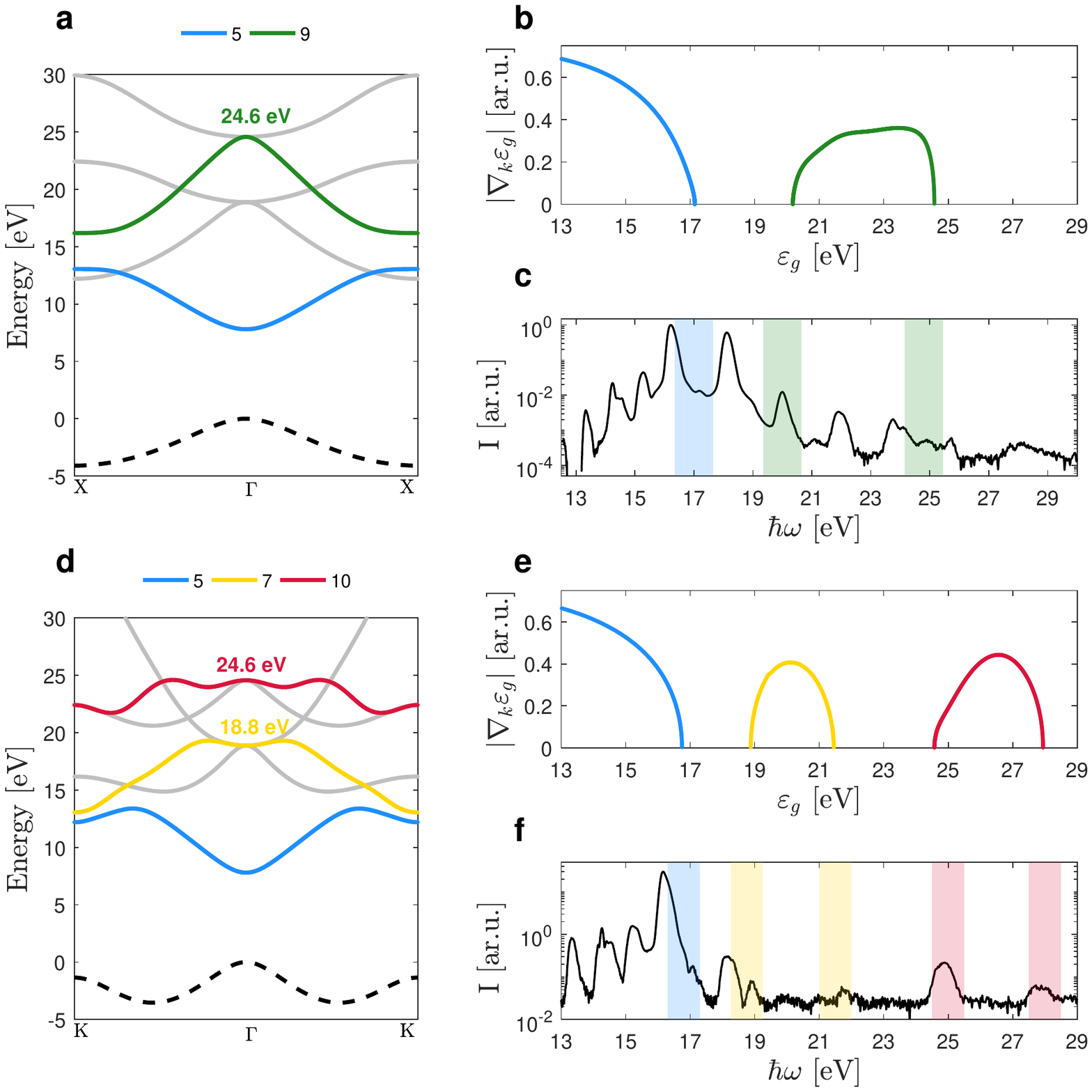}
\caption{\textbf{Spectral caustics at singular points of the dynamical JDOS}. The band structure along the $0^\circ$ ($\Gamma-X$) (\textbf{a}) and $45^\circ$ ($\Gamma-K$) (\textbf{d}) orientations, respectively. The dashed line is the valence band and the colored lines are the conduction bands which are dipole coupled to the valence band. The gray bands indicate the existing bands which are not dipole coupled. \textbf{b} and \textbf{e} plot $\nabla_{\textbf{k}}\varepsilon_g(\textbf{k})$ along $0^\circ$ and $45^\circ$, respectively. The energies for which $\nabla_{\textbf{k}}\varepsilon_g(\textbf{k})=0$, where singularities in the dynamical JDOS occur, are highlighted by corresponding colored bars on the oscillating spectra \textbf{c} ($0^\circ$) and \textbf{f} ($45^\circ$), emphasizing their link to the spectral caustics.}
\label{fig:fig2}
\end{figure}

We start by analyzing the enhancement at 23.5-26 eV. This observation is striking --  clearly, this spectral region originates from a high conduction band excitation associated with a low population transfer. Furthermore, while the oscillating spectrum changes dramatically with the crystal orientation, this feature remains robust. Such observation suggests that it originates from areas of the BZ which are orientation independent. These areas can be found around the $\Gamma$ point of the BZ which is common to all crystal orientations. Figures \ref{fig:fig2}\textbf{a} and \ref{fig:fig2}\textbf{b} describe the band structure for  crystal orientations of $0^\circ$ and $45^\circ$, respectively. Indeed, we find two high conduction bands intersecting the $\Gamma$ point at an energy of 24.57 eV. At the $\Gamma$ point these bands, as well as any band, have a critical point, $\nabla_{\textbf{k}}E_c(\textbf{k})=\nabla_{\textbf{k}}E_v(\textbf{k})=0$, such that the gradient of the energy gap is zero as well:
 \begin{equation}\label{caustic condition}
 \nabla_{\textbf{k}}\left(E_c(\textbf{k})-E_v(\textbf{k})\right)=\nabla_{\textbf{k}}\varepsilon_g(\textbf{k})=0
  \end{equation}

where $\textbf{k}$ is the lattice momentum, $E_c$ is the conduction band and $E_v$ is the valence band. At the critical points, where $\nabla_{\textbf{k}}\varepsilon_g(\textbf{k})=0$, the JDOS of a crystalline solid becomes singular. These singularities are known as Van Hove singularities and most commonly arise in the analysis of optical absorption and reflection spectra \cite{ehrenreich1962optical,roessler1967electronic}. In contrast to these measurements, performed over long time scales, HHG in bulk crystal is attributed to the sub-cycle electronic dynamics where the electron and hole wave-packets remain mutually coherent. In addition, these wave-packets, which are initiated by tunneling, are localized in the BZ. As a result, the harmonic emission can not be understood by merely counting the total number of available states throughout the BZ. Instead, the quantum nature of the singularity has to be considered at specific \textbf{k} regions. During such short time scales, we can view the singularities as points where the relative velocity of the electron and hole wave-packets is zero. This results in enhanced interference effects, encoding dynamical quantum information into the HHG spectrum.

  The quantum expression for the interband currents can be written as \cite{vampa2014theoretical,vampa2015semiclassical}

 \begin{gather}\label{interband current}
 \textbf{j}_{er}(\omega)=\omega\int_{B.Z.}d^3\textbf{k}\iint \textbf{g}(\textbf{k},t',t)\exp^{-iS(\textbf{k},t',t)+i\omega t}dtdt'\\
 S(\textbf{k},t',t)=\int_{t'}^{t}\varepsilon_g(\textbf{k}-\textbf{A}(t)+\textbf{A}(\tau))d\tau
   \end{gather}
where $\textbf{A}$ is the vector potential and $\omega$ the harmonic energy. $g(\textbf{k},t',t)$ represents the slowly varying term compared to the fast oscillating exponential containing the semi-classical action, $S(\textbf{k},t',t)$. Under these conditions we can use the stationary phase approximation (SPA) for $\textbf{j}_{er}(\omega)$ which defines the semi-classical mapping between harmonic energy and recolliding electron and hole trajectories. The interband current for a single stationary solution as obtained from the SPA, is proportional to $\sqrt{|\hat{\textbf{S}}''(\textbf{k}_{st},t'_{st},t_{st})|}^{-1}$, where $\hat{\textbf{S}}''(\textbf{k}_{st},t'_{st},t_{st})$ is the Hessian matrix  of the function S(\textbf{k},t',t) at the stationary points. This derivation directly links the HHG spectrum and the band structure (for a detailed derivation see SI):
\begin{equation}\label{spectral_intensity}
    I(\omega)\propto|\textbf{j}_{er}(\omega)|^2\propto\omega^2\left|\sum_{\textbf{k}_{st}}\frac{\tilde{\textbf{g}}(\textbf{k}_{st},t'_{st},t_{st})e^{-iS(\textbf{k}_{st},t'_{st},t_{st})}}{\sqrt{|\nabla_{\textbf{k}}\varepsilon_{g}(\textbf{k}_{st})|}}\right|^2
\end{equation}
where $\varepsilon_g(\textbf{k}_{st})$ is the energy difference between electron and hole at the time of recollision, which defines the harmonic energy, and $\tilde{\textbf{g}}$ accounts for all pre-exponential terms. The expression for the spectral intensity strongly resembles that of the JDOS, except it is weighted by the exponent of the semi-classical action, therefore, it is associated with a \textit{dynamical JDOS}. This exponent gives the quantum phase and amplitude associated with each trajectory. Due to strong field tunneling, the quantum amplitude strongly attenuates with increasing $|\textbf{k}_{st}^\perp|$, $\textbf{k}_{st}^\perp$ being the component of $\textbf{k}_{st}$ perpendicular to the laser's polarization. Such attenuation expresses the fact that HHG indeed originates from a narrow stripe of the BZ, providing the significant angular dependence of the spectrum. Clearly this mapping becomes singular at the extrema of the energy gap as described by equation \eqref{caustic condition}.

The singularity in the mapping between harmonic energy and electron-hole trajectories can be described within the framework of caustics \cite{raz2012spectral}. Caustics are universal phenomena in nature that link processes observed in many different branches of physics. Previous studies identified the appearance of caustics in gas phase HHG \cite{raz2012spectral,facciala2016probe}, where they reveal the quantum nature of the process in a regime where classical analysis fails. Figure \ref{fig:fig1}\textbf{e} illustrates the origin of spectral caustics in condensed matter systems, when the electron-hole wave-packet has zero relative velocity at the extrema of the band gap. We find that spectral caustics reveal the rich, quantum,  spatio-temporal nature of electronic dynamics in solids, dictated by the direct link between the band structure and strong field interaction.

The enhancement at 23.5-26 eV can be identified as a spectral caustic originating from the $\Gamma$ point. Figures \ref{fig:fig2}\textbf{a}, \ref{fig:fig2}\textbf{d} present the band structure for crystal orientations of $0^\circ$ and $45^\circ$, respectively. Another energy band intersecting the $\Gamma$ point can be found at 18.8 eV (figure \ref{fig:fig2}\textbf{d}), leading to a robust spectral feature near 18 eV as can be seen in figures \ref{fig:fig1}\textbf{a}-\textbf{c}. While the crossing of energy bands at the $\Gamma$ point will always result in singularities, spectral caustics can be found at other points in the BZ as well. Along the high symmetry axes of the crystal, $0^\circ$ and $45^\circ$, each band-gap consists of several critical points where $\nabla_{\textbf{k}}\varepsilon_g(\textbf{k}_{st})=0$. In figure \ref{fig:fig2} we show how these points are directly imprinted in the experimentally resolved HHG spectrum. Figures \ref{fig:fig2}\textbf{b}, \ref{fig:fig2}\textbf{e} describe the gradient of each band gap  as a function of the band gap energy, for crystal orientations of $0^\circ$ and $45^\circ$, respectively. Figures \ref{fig:fig2}\textbf{c}, \ref{fig:fig2}\textbf{f} present the measured oscillating spectra at these angles. The highlighted lines in figures \ref{fig:fig2}\textbf{c}, \ref{fig:fig2}\textbf{f} emphasize  the energy points where the gradient is zero and the dynamical JDOS becomes singular according to equation \eqref{spectral_intensity} . Indeed, caustics dominate the brightest features in the HHG spectrum, leading to a dramatic spectral focusing and enhancement of the weak signal associated with higher conduction bands excitation.

\begin{figure}[h]
\centering
\includegraphics[trim=45 135 25 125,clip,width=0.95\linewidth]{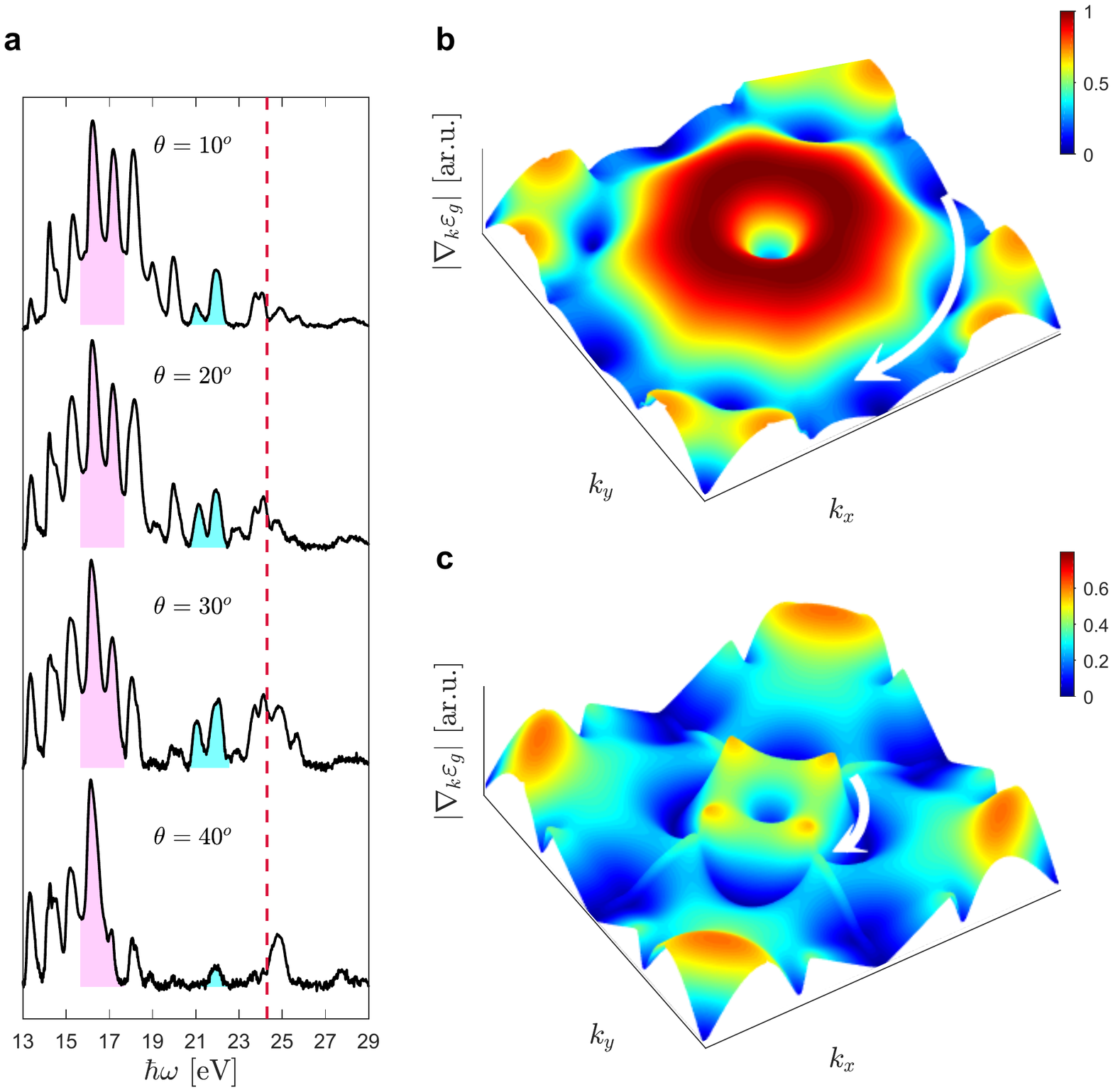}
\caption{\textbf{HHG spectral enhancement at high dynamical JDOS valleys}. \textbf{a} the oscillating spectrum for $\theta=10^\circ,20^\circ,30^\circ,40^\circ$. Two enhanced spectral structures which persist over many orientations are marked by violet and cyan. These spectral structures originate from high dynamical JDOS. The dashed red line marks the spectral caustic between 23.5-26 eV highlighting its angular dependence.  \textbf{b} and \textbf{c} are the 2D gradients of the band gap for bands 5 and 7, respectively. The arrows show valleys of low $\nabla_{\textbf{k}}\varepsilon_g(\textbf{k})$, i.e. high dynamical JDOS, corresponding to the enhanced spectral features in \textbf{a}. The valley in \textbf{b}  corresponds to the structure marked by violet and \textbf{c} to the one in cyan. }
\label{fig:fig3}
\end{figure}

While the singularities are observed along the high symmetry axes of the crystal, figure \ref{fig:fig3}\textbf{a} shows that enhanced spectral features are observed at other angles as well . Their origin can be understood through the dynamical JDOS, by looking at the gradient of the different band gaps, $\nabla_{\textbf{k}}\varepsilon_g(\textbf{k})$. Figures \ref{fig:fig3}\textbf{b} and \ref{fig:fig3}\textbf{c} present 2D images of the gradients of bands 5 and 7. In both bands we can identify a pronounced valley at specific $\textbf{k}$ values and across a large angular range. Along these valleys $\nabla_{\textbf{k}}\varepsilon_g(\textbf{k})$ is small, leading, according to equation \eqref{spectral_intensity} to a local spectral enhancement in the HHG signal. Specifically, the gradient valley at band 5 is mapped to a pronounced spectral enhancement around 16-17 eV, while the valley in band 7 leads to the enhanced spectral feature in the 20.5-22.5 eV region, marked by the violet and cyan areas, respectively. Importantly, their observation is robust with the crystal orientation, mapping the angular variation of the gradient valleys.

Understanding the mapping between the band structure and the HHG spectrum allows us to study the fundamental dynamical properties of strong field dressed phenomena in solids \cite{hawkins2015effect,mcdonald2017intense}. Such effects play an important role when the band structure becomes degenerate. In the vicinity of these regions the intricate interplay between the strong field dressing and the orientation dependent band couplings can have a dramatic effect on the electronic dynamics. At the $\Gamma$ point of MgO there is a two band degeneracy at 24.6 eV (bands 9 and 10, plotted in green and in red), protected by the crystal symmetries. Taking a closer look at the associated caustic we find that our measurement encodes a signature of such a strong field mechanism. The enhancement observed at the high energy end of the spectrum, 23.5-26 eV, is robust with crystal orientation, as one would expect from a feature originating from the $\Gamma$ point. However, the spectral shape of this feature shows a striking angular dependence, where the peak shifts from harmonic 25 at $0^\circ$ to 26 at $45^\circ$, exhibiting complex spectral structures at intermediate angles (see dashed line in figure \ref{fig:fig3}\textbf{a}). By rotating the crystal we change the couplings and field dressing of these two bands, imprinting the intricate underlying strong field dynamics into the angular dependence of the HHG spectrum.

Our results establish the fundamental connection between the electronic structure of the crystal and the strong field process. This study reveals how strong field attosecond metrology serves as an extremely sensitive probe of ultrafast dynamical quantum interference between electron-hole wave-packets in solids. We identify the important role of the dynamical JDOS, dictated by the strong field nature of the interaction. As the dynamical JDOS becomes singular, the trajectory picture that governs the interband HHG process fails, giving rise to spectral caustics. These findings provide a unique insight into the dressing of the band structure by the strong laser field. The mapping between the HHG spectrum and the band gaps serves as unequivocal evidence to the dominant role of interband emission over numerous conduction bands. Looking forward, our study will form the framework for a large range of attosecond scale phenomena, opening a new path in the study of quasi-particle interactions within the field dressed band structure of crystals. It will allow the study of ultrafast dynamics such as electron-hole interactions, leading to the formation of excitons, or electron-electron-phonon interactions. Furthermore, the control over sub-cycle electronic currents will play a role in the establishment of compact solid state XUV sources, as well as in the field of petahertz electronics.

\section*{Data Avilability}
The data that support the findings of this study are available from the corresponding author upon reasonable request.

\bibliographystyle{naturemag}
\bibliography{bib_MgO}

\end{document}


\title{Supplementary Information}
\author[1]{Ayelet Julie Uzan\footnote{These authors contributed equally to this work}}
\author[1]{Gal Orenstein$^*$}
\author[2]{\'Alvaro Jim\'enez-Gal\'an}
\author[3]{Chris McDonald}
\author[4,2]{Rui E.F Silva}
\author[1]{Barry D. Bruner}
\author[5, 6]{Nikolai D. Klimkin}
\author[7]{Valerie Blanchet}
\author[1]{Talya Arusi-Parpar}
\author[1]{Michael Kr\"uger}
\author[5, 6]{Alexey N. Rubtsov}
\author[2,8]{Olga Smirnova}
\author[2,9,10]{Misha Ivanov}
\author[1]{Binghai Yan}
\author[3]{Thomas Brabec}
\author[1]{Nirit Dudovich}
\affil[1]{\footnotesize Department of Complex Systems, Weizmann Institute of Science, 76100, Rehovot, Israel}
\affil[2]{\footnotesize Max-Born-Institut, Max-Born Strasse 2A, D-12489 Berlin, Germany}
\affil[3]{\footnotesize Department of Physics, University of Ottawa, Ottawa, Ontario K1N 6N5, Canada}
\affil[4]{\footnotesize Department of Theoretical Condensed Matter Physics, Universidad Autónoma de Madrid, E-28049 Madrid, Spain}
\affil[5]{\footnotesize Russian Quantum Center, Skolkovo 143025, Russia}
\affil[6]{\footnotesize Department of Physics, Moscow State University, 119991 Moscow, Russia}
\affil[7]{\footnotesize Universit\'e  de  Bordeaux -CNRS,  CELIA, UMR5107,  F33405  Talence,  France}
\affil[8]{\footnotesize Technische Universit\"at Berlin, Ernst-Ruska-Geba\"ude, Hardenbergstr. 36A, D-10623 Berlin, Germany}
\affil[9]{\footnotesize Blackett Laboratory, Imperial College London, South Kensington Campus, SW7 2AZ London, United Kingdom}
\affil[10]{\footnotesize Department of Physics, Humboldt University, Newtonstrasse 15, 12489 Berlin, Germany}

\renewcommand\Authands{ and }
\date{\today}%
\maketitle 

\section{Experimental set up}
High harmonics are generated with a near-infrared laser source (Light Conversion Topas-HE) centered around $1.3 \mu m$ wavelength at 1kHz repetition rate and pulse duration of $50 fs$. The beam is focused into a free standing single crystal of MgO ($<100>$) with $100 \mu m$ thickness. The intensity of the beam at the focus is $1.5\cdot10^{13} \frac{W}{cm^2}$ which is just under the damage threshold of MgO. The orientation of the crystal axes with respect to the laser polarization is controlled by a rotation motor perpendicular to the beam propagation direction ($\theta=0$ is determined by the HHG efficiency). The sample is placed inside the vacuum chamber at the imaging plane of the XUV grating.
The second harmonic field, polarized perpendicular to the fundamental field, is generated using a $100 \mu m$ type-$I$ beta barium borate (BBO) ($\beta BaB_{2}O_4$) crystal. Group-velocity dispersion is compensated using a birefringent crystal (calcite). The subcycle delay of the SH relative to the fundamental field is controlled using a pair of fused silica wedges. We record the harmonic spectra with an imaging spectrometer consisting of a flat-field variable groove density grating and microchannel plates.

\section{Two color as lock-in measurement}
\begin{figure}[h]
\centering
\includegraphics[trim= 175 330 90 200,clip,width=1\linewidth]{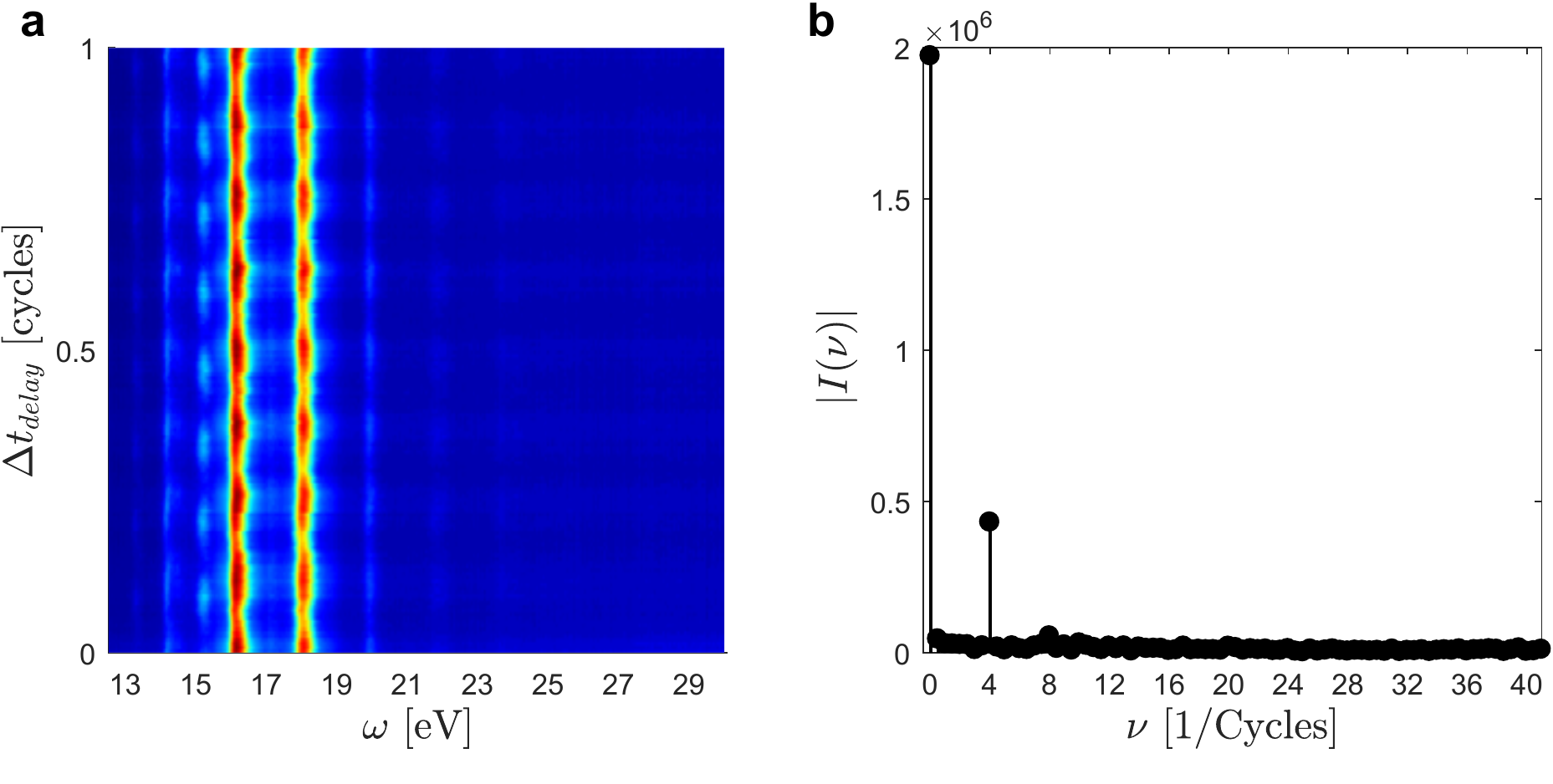}
\caption{\textbf{a} Two color oscillations of the harmonic spectrum at $0^\circ$ orientation of the crystal. \textbf{b} The amplitude of the Fourier transform of harmonic 19, presents a peak at $\nu=4$ and a DC component.}
\label{fig:sifig1}
\end{figure}

Performing two color measurements enables us to detect very weak signals, mainly the harmonics' emission in the high energy range of the spectrum. Since these harmonics originate from high conduction bands, the ability to detect their signal and angular dependence is a crucial step in recording the multi-band electron dynamics.
In analogy to lock-in measurement methods, periodic modulation of the harmonics signal allows the separation of the oscillating signal from the DC background by applying a Fourier analysis. Modulation of the harmonics signal can be achieved by engineering the instantaneous driving field. Adding the second harmonic field and scanning the delay between the two fields, modulates the harmonics signal at four times the fundamental frequency (see Figure \ref{fig:sifig1}\textbf{a}). We Fourier transform the oscillating harmonic signal, along the two color delay. In Figure \ref{fig:sifig1}\textbf{b} we present the absolute value of the Fourier transform of harmonic 19 (measured at $0^\circ$ orientation angle of the crystal). As was presented in the main text (main text figures 1a-1c), by extracting the Fourier amplitude at $\nu=4$, for each harmonic in the spectrum, we increase the SNR-- detecting very weak signals that could barley be observed in the single color spectrum.

\section{Higher Conduction Bands Stationary Phase Approximation}
The stationary phase approximation (SPA) is a common approach to evaluate strong field light matter interaction, as the interband current in solids. However, the expression in the main text (eq. 2) is valid only for the case of one conduction band. In the following we will develop a similar expression for the interband current arising from a higher conduction band, considering a two conduction band scenario. This analysis can be generalized to a larger number of bands. We begin by writing the three band semi-conductor Bloch equations \cite{mcdonald2015interband}. The valence band is indexed as 1 and the conduction bands are 2 and 3:
\begin{gather}
n_m=|b_m|^2 \quad \pi_{mm'}=b^*_{m}b_{m'} \label{population}\\
S_{mm'}(\textbf{K},t)=\int_{-\infty}^{t}\varepsilon_{mm'}(\textbf{K}+\textbf{A}(\tau))d\tau \quad \varepsilon_{mm'}(\textbf{k})=E_m(\textbf{k})-E_{m'}(\textbf{k}) \\
\dot{n}_m=i\sum_{m'\neq m}\Omega_{mm'}\pi_{mm'}e^{iS_mm'}+c.c. \\
\dot{\pi}_{12}=i \Omega^*_{12}(n_1-n_2)e^{-iS_{12}}+i\Omega_{23}\pi_{13}e^{iS_{23}}-i\Omega^*_{13}\pi^*_{23}e^{-iS_{13}} \label{pi12}\\
\dot{\pi}_{13}=i \Omega^*_{13}(n_1-n_3)e^{-iS_{13}}+i\Omega_{32}\pi_{12}e^{iS_{32}}-i\Omega^*_{12}\pi^*_{32}e^{-iS_{12}} \label{pi13}\\
\dot{\pi}_{23}=i \Omega^*_{23}(n_2-n_3)e^{-iS_{23}}+i\Omega_{31}\pi_{21}e^{iS_{31}}-i\Omega^*_{21}\pi^*_{31}e^{-iS_{21}} \label{pi23}
\end{gather}
where $n_m$ and $\pi_{mm'}$ are the population in the m band and the coherence between m and m' bands respectively, \textbf{K}=\textbf{k}-\textbf{A}(t), \textbf{A}(t) is the vector potential, $E_m(\textbf{k})$ is the energy of the m band, $\Omega_{mm'}=\textbf{F}\cdot\textbf{d}_{mm'}$, $\textbf{F}=-\frac{d\textbf{A}}{dt}$ and $\textbf{d}_{mm'}$ is the transition dipole moment between bands m and m'. For a full derivation see \cite{mcdonald2015interband}. The interband polarization between bands m and m' is given by:
\begin{equation}
\textbf{p}_{mm'}(\textbf{K},t)=\textbf{d}_{mm'}[\textbf{K}+\textbf{A}(t)]\pi_{mm'}(\textbf{K},t)e^{iS_{mm'}(\textbf{K},t)}+c.c.
\end{equation}

We are interested in the polarization of the higher band (band 3) and therefore must solve the equation for $\pi_{13}$. To do so we assume that the population of the bands drops rapidly as we go to higher conduction bands i.e. $n_m>>n_{m'}$ for $m'>m$ and $n_1=1$. Using equation \eqref{population} we derive a stronger form of this condition:
\begin{equation}
|b_1|>>|b_{2}|>>|b_3|
\end{equation}
As with the SPA, these approximations do not necessarily limit the validity of our analysis. They are simply used to get an analytical solution to an otherwise complicated system of equations, which will allow us to directly explore the effects of the band structure. Applying these approximations to equations \eqref{pi13} and \eqref{pi12} yields:
\begin{gather}
\dot{\pi}_{12}=i\Omega^*_{12}e^{-iS_{12}} \label{pi12approx}\\
\dot{\pi}_{13}=i \Omega^*_{13}e^{-iS_{13}}+i\Omega_{32}\pi_{12}e^{iS_{32}} \label{pi13approx}
\end{gather}
These equations can now be directly integrated. We see that $\pi_{13}$ consists of two terms. The first term in equation \eqref{pi13approx} will give a similar solution to a two band system which we have already presented in the main text. This solution may be strongly suppressed since it involves tunneling through a very high energy barrier. Integrating equations \eqref{pi12approx} and \eqref{pi13approx}, considering only the second term in \eqref{pi13approx} gives the desired solution for the interband current:
\begin{gather}
\begin{gathered}\label{multiband_current}
\textbf{j}_{er}(\omega)=\omega\int_{B.Z.}d^3\textbf{k}\int dt_3\textbf{p}_{mm'}(\textbf{k},t_3)e^{i\omega t_3}=\\
=\omega\int_{B.Z.}d^3\textbf{k}\iiint \textbf{g}(\textbf{k},t_1,t_2,t_3)\exp^{-iS(\textbf{k},t_1,t_2,t_3)+i\omega t_3}dt_1dt_2dt_3
\end{gathered} \\
S(\textbf{k},t_1,t_2,t_3)=\int_{t_1}^{t_2}\varepsilon_{g1}(\textbf{k}-\textbf{A}(t_3)+\textbf{A}(\tau))d\tau+\int_{t_2}^{t_3}\varepsilon_{g2}(\textbf{k}-\textbf{A}(t_3)+\textbf{A}(\tau))d\tau \label{3bandS}\\
\varepsilon_{gn}(\textbf{k})=E_{n+1}(\textbf{k})-E_{1}(\textbf{k})
\end{gather}
$g(\textbf{k},t_1,t_2,t_3)$ is a slowly varying term compared to the fast oscillating exponential containing $S(\textbf{k},t_1,t_2,t_3)$. Since band 1 is the valence band and 2,3 are the conduction bands we get a very similar expression to the interband current (equation 2) in the main text. This analysis can be generalized to $N$ conduction bands giving the following exponential term:
\begin{equation}\label{multiband_S}
S(\textbf{k},t_1,...,t_{N+1})=\sum_{n=1}^{N}\int_{t_n}^{t_{n+1}}\varepsilon_{gn}(\textbf{k}-\textbf{A}(t_{N+1})+\textbf{A}(\tau))d\tau
\end{equation}

where $n$ represents the conduction band number.
\section{Derivation of the connection between interband currents and the band structure}
We approximate the integral of the multi-band interband current (equations \eqref{multiband_current},\eqref{multiband_S}) using the SPA. Essentially we assume that the exponential term oscillates very fast so only points of relatively stationary phase will contribute. We do this by expanding $S(\textbf{k},t_r,t_1,...,t_{N})$ to second order requiring that the first order is 0. For abbreviation we use $t_{N+1}=t_r$ and the corresponding stationary time and energy gap are $t_{(N+1)st}=t_{(r)st}$ and $\varepsilon_{g(N+1)}=\varepsilon_{g(r)}$ we also abbreviate $t_1,...,t_{N}$ to $t_{1:N}$. This gives us the stationary equations:
\begin{gather}\label{eq_stat_1}
\nabla_{\textbf{k}}S(\textbf{k}_{st},t_{(r)st},t_{1:Nst})=0 \\\label{eq_stat_2}
\frac{\partial S(\textbf{k}_{st},t_{(r)st},t_{1:Nst})}{\partial t_{1:N}}=0\\
\begin{gathered}
\frac{\partial\label{eq_stat_3} S(\textbf{k}_{st},t_{(r)st},t_{1:Nst})}{\partial t_{r}}=\\
=\varepsilon_{g(r)}(\textbf{k}_{st})-\dot{\textbf{A}}(t_{(r)st})\cdot\nabla_{\textbf{k}}S(\textbf{k}_{st},t_{(r)st},t_{1:Nst})-\omega=\\
=\varepsilon_{g(r)}(\textbf{k}_{st})-\omega=0
\end{gathered}
\end{gather}
Expanding $S$ around the stationary points gives:
\begin{equation}
S\approx S_0+\textbf{v}_\textbf{st}\hat{\textbf{S}}''\textbf{v}_\textbf{st}^T
\end{equation}
Where $\hat{\textbf{S}}''$ is the Hessian matrix of $S$ and $\textbf{v}_\textbf{st}=(\textbf{k}_{st},t_{(r)st},t_{1:Nst})$. Plugging this back into the interband current integral (equation \eqref{multiband_current}), we get a multidimensional Gaussian integral. Its solution is proportional to $\sqrt{|\hat{\textbf{S}}''|^{-1}}$, where $|\hat{\textbf{S}}''|$ is the determinant of the Hessian (we use the stationary equations \eqref{eq_stat_1}, \eqref{eq_stat_2}, \eqref{eq_stat_3} to simplify expressions):

\begin{gather}\label{hessian}
\begin{gathered}
|\hat{\textbf{S}}''|=\\
\left|
\begin{array}{ccccc}
\partial_{xx}S  & \partial_{yx}S  & \partial_{zx}S  & \partial_{t_{r}x}S & \partial_{t_{1:N}x}S    \\
\partial_{xy}S  & \partial_{yy}S & \partial_{zy}S & \partial_{t_{r}y}S & \partial_{t_{1:N}y}S     \\
\partial_{xz}S  & \partial_{yz}S  & \partial_{zz}S & \partial_{t_{r}z}S  & \partial_{t_{1:N}z}S    \\
\left(\begin{gathered}
\partial_x\varepsilon_{g(r)}+\\ \dot{\textbf{A}}(t_{(r)st})\cdot \\\partial_x\nabla_{\textbf{k}}S
\end{gathered}\right)
& \left(\begin{gathered}
\partial_y\varepsilon_{g(r)}+\\ \dot{\textbf{A}}(t_{(r)st})\cdot \\\partial_y\nabla_{\textbf{k}}S
\end{gathered}\right)  & \left(\begin{gathered}
\partial_z\varepsilon_{g(r)}+\\ \dot{\textbf{A}}(t_{(r)st})\cdot \\\partial_z\nabla_{\textbf{k}}S
\end{gathered}\right) &  \left(\begin{gathered}
\dot{\textbf{A}}(t_{(r)st})\cdot \\\partial_{t_{r}}\nabla_{\textbf{k}}S
\end{gathered}\right) & \left(\begin{gathered}
\dot{\textbf{A}}(t_{(r)st})\cdot \\\partial_{t_{1:N}}\nabla_{\textbf{k}}S
\end{gathered}\right)    \\
\partial_{xt_{1:N}}S  & \partial_{yt_{1:N}}S  & \partial_{zt_{1:N}}S & \partial_{t_{r}t_{1:N}}S & \partial_{t_{1:N}t_{1:N}}S    \\
\end{array}
\right|=
\\
\left|\hat{\textbf{M}}_1\right|+\left|\hat{\textbf{M}}_2\right|
\end{gathered}
\end{gather}
Where $\partial_x$, $\partial_y$, $\partial_z$ denote the partial derivatives with respect to $k_x$, $k_y$ and $k_z$ respectively. The last row and column of this matrix extend to N rows and columns over all corresponding times.
\begin{equation}\label{m1}
\begin{gathered}
\left|\hat{\textbf{M}}_1\right|=\\
\left|
\begin{array}{ccccc}
\partial_{xx}S  & \partial_{yx}S  & \partial_{zx}S  & \partial_{t_{r}x}S & \partial_{t_{1:N}x}S    \\
\partial_{xy}S  & \partial_{yy}S & \partial_{zy}S & \partial_{t_{r}y}S & \partial_{t_{1:N}y}S     \\
\partial_{xz}S  & \partial_{yz}S  & \partial_{zz}S & \partial_{t_{r}z}S  & \partial_{t_{1:N}z}S    \\
\partial_x\varepsilon_{g(r)}  &  \partial_y\varepsilon_{g(r)}  &
\partial_z\varepsilon_{g(r)}   &  0 & 0  \\
\partial_{xt_{1:N}}S  & \partial_{yt_{1:N}}S  & \partial_{zt_{1:N}}S & \partial_{t_{r}t_{1:N}}S & \partial_{t_{1:N}t_{1:N}}S    \\
\end{array}
\right|=\\
=\nabla_{\textbf{k}}\varepsilon_{g(r)}(\textbf{k}_{st})\cdot\textbf{f}(\textbf{k}_{st},t_{(r)st},t_{1:Nst})
\end{gathered}
\end{equation}
where $\textbf{f}(\textbf{k}_{st},t_{(r)st},t_{1:Nst})$ is a vector given by the relevant minors of $\hat{\textbf{M}}_1$.
\begin{equation}\label{m2}
\begin{gathered}
\left|\hat{\textbf{M}}_2\right|=\\
\left|
\begin{array}{ccccc}
\partial_{xx}S  & \partial_{yx}S  & \partial_{zx}S  & \partial_{t_{r}x}S & \partial_{t_{1:N}x}S    \\
\partial_{xy}S  & \partial_{yy}S & \partial_{zy}S & \partial_{t_{r}y}S & \partial_{t_{1:N}y}S     \\
\partial_{xz}S  & \partial_{yz}S  & \partial_{zz}S & \partial_{t_{r}z}S  & \partial_{t_{1:N}z}S    \\
\left(\begin{gathered}
\dot{\textbf{A}}(t_{(r)st})\cdot \\\partial_x\nabla_{\textbf{k}}S
\end{gathered}\right)
& \left(\begin{gathered}
\dot{\textbf{A}}(t_{(r)st})\cdot \\\partial_y\nabla_{\textbf{k}}S
\end{gathered}\right)  & \left(\begin{gathered}
\dot{\textbf{A}}(t_{(r)st})\cdot \\\partial_z\nabla_{\textbf{k}}S
\end{gathered}\right) &  \left(\begin{gathered}
\dot{\textbf{A}}(t_{(r)st})\cdot \\\partial_{t_{r}}\nabla_{\textbf{k}}S
\end{gathered}\right) & \left(\begin{gathered}
\dot{\textbf{A}}(t_{(r)st})\cdot \\\partial_{t_{1:N}}\nabla_{\textbf{k}}S
\end{gathered}\right)    \\
\partial_{xt_{1:N}}S  & \partial_{yt_{1:N}}S  & \partial_{zt_{1:N}}S & \partial_{t_{r}t_{1:N}}S & \partial_{t_{1:N}t_{1:N}}S    \\
\end{array}
\right|
\end{gathered}
\end{equation}
It is clear from \eqref{m2} that the rows of $\hat{\textbf{M}}_2$ are linearly dependent:
\begin{equation}
\dot{A}_x(t_{(r)st})\hat{\textbf{M}}_2(1,j)+\dot{A}_y(t_{(r)st})\hat{\textbf{M}}_2(2,j)+\dot{A}_z(t_{(r)st})\hat{\textbf{M}}_2(3,j)=\hat{\textbf{M}}_2(4,j)
\end{equation}
Where $\hat{\textbf{M}}_2(i,j)$ is the element in the $i$ row and $j$ column of $\hat{\textbf{M}}_2$ and $\textbf{A}=(A_x,A_y,A_z)$. As a result $\left|\hat{\textbf{M}}_2\right|=0$. Finally from equations \eqref{hessian},\eqref{m1},\eqref{m2} we derive the interband current:
\begin{equation}\label{stationary_j}
\begin{gathered}
\textbf{j}_{er}(\omega)\propto\omega\sum_{
\substack{\text{stationary}\\\text{solutions}}}\frac{\textbf{g}(\textbf{v}_\textbf{st})e^{-iS(\textbf{v}_\textbf{st})}}{\sqrt{\nabla_{\textbf{k}}\varepsilon_{g(r)}(\textbf{k}_{st})\cdot{\textbf{f}}(\textbf{v}_\textbf{st})}}
\end{gathered}
\end{equation}
Equation \eqref{stationary_j} establishes the connection between the interband current and the gradient of the band structure.

\section{Dynamical JDOS}

The SPA of the interband current (equation \eqref{stationary_j}) is based on the summation over all possible stationary solutions. We divide these solutions into $\textbf{k}_{st}$ families. Each family is composed of solutions with the same $\textbf{k}_{st}$ but different stationary times. These solutions recombine at the same point in the Brillouin zone, but take different trajectories to get there. For example long and short trajectories are two members of this family of solutions. Equation \eqref{eq_stat_3} suggests that all $\textbf{k}_{st}$'s lying on the surface $\varepsilon_{g(r)}(\textbf{k}_{st})=\omega$ are possible solutions. For gas phase harmonics, equation \eqref{eq_stat_1} requires \textbf{k}$_{st}^\perp$=0, where \textbf{k}$_{st}^\perp$ is the component of  \textbf{k}$_{st}$ perpendicular to the laser polarization. This gives a single possible  \textbf{k}$_{st}$, meaning that there is only one $\textbf{k}_{st}$ family. However, in the general case, which applies to MgO, the band gap is not separable in $\text{k}_x$, $\text{k}_y$, $\text{k}_z$ (i.e.there is no rotation which allows us to write $\varepsilon_{g(r)}(\textbf{k})=\varepsilon_{x}(\text{k}_x)+\varepsilon_{y}(\text{k}_y)+\varepsilon_{z}(\text{k}_z)$) there are many more possible solutions for $\textbf{k}_{st}$. Defining $\nabla_{\textbf{k}}\varepsilon_{g(r)}(\textbf{k})=|\nabla_{\textbf{k}}\varepsilon_{g(r)}(\textbf{k})|\hat{\textbf{n}}(\textbf{k})$ and $\tilde{\textbf{g}}(\textbf{v}_\textbf{st})=\frac{\textbf{g}(\textbf{v}_\textbf{st})}{\sqrt{\hat{\textbf{n}}(\textbf{k}_{st})\cdot\textbf{f}(\textbf{v}_\textbf{st})}}$ we can write an expression for the spectral intensity of the harmonics:
\begin{equation}\label{spectral_inensity}
 I(\omega)\propto|\textbf{j}_{er}(\omega)|^2\propto\omega^2\left|\sum_{\textbf{k}_{st}}\sum_{\text{Trajectories}}\frac{\tilde{\textbf{g}}(\textbf{v}_\textbf{st})e^{-iS(\textbf{v}_\textbf{st})}}{\sqrt{|\nabla_{\textbf{k}}\varepsilon_{g(r)}(\textbf{k}_{st})|}}\right|^2
\end{equation}
We now focus on a single trajectory type, which is often the correct picture for HHG as short trajectories tend to dominate the spectrum. This solution is described by equation (4) in the main text. We compare this expression to the JDOS:
\begin{equation}\label{jdos}
g(\text{E})=\oiint\limits_{\varepsilon_{g}(\textbf{k})=\text{E}}\frac{dS_\text{k}}{4\pi^3}\frac{1}{|\nabla_{\textbf{k}}\varepsilon_{g}(\textbf{k})|}
\end{equation}

Keeping in mind that the  $\textbf{k}_{st}$ summation in \eqref{spectral_inensity} is over a constant energy surface similar to the integral in \eqref{jdos} we see that these expressions are very close. In fact if we set $\tilde{\textbf{g}}(\textbf{v}_\textbf{st})e^{-iS(\textbf{v}_\textbf{st})}$=const we get:
\begin{equation}
 \tilde{I}(\omega)\propto\omega^2\sum_{\textbf{k}_{st}}\frac{1}{|\nabla_{\textbf{k}}\varepsilon_{g(r)}(\textbf{k}_{st})|}
\end{equation}
$\tilde{\textbf{g}}(\textbf{v}_\textbf{st})$ varies slowly with the stationary solutions, however, $e^{-iS(\textbf{v}_\textbf{st})}$ can change quite rapidly. Since this term accounts for tunneling, which is strongly suppressed with increasing band gap, the exponent decays rapidly with increasing $|\textbf{k}_{st}^\perp|$. As a result the spectral intensity samples the JDOS in a narrow region of the Brillouin zone along the laser polarization, hence the term dynamical JDOS. This means that unlike many other optical measurements such as absorption and reflection which sample the entire Brillouin zone, the narrow electronic wave-packet initiated by tunneling is sensitive to a smaller region of the Brillouin zone. This is the origin of the angular resolution and features in the HHG spectrum which are strongly related to the localization of the JDOS in crystal momentum space. 

\section{Mapping between crystal momentum and harmonic energy}

In the interband picture, the associated crystal momentum of each recombining electron-hole pair is dictated by energy conservation – $\varepsilon_g(\textbf{k})=\omega$. This relation provides the mapping between crystal momentum and the HHG spectrum. In the main text we used this mapping to describe the harmonic enhancement determined by the gradient of the gap which is calculated in \textbf{k} space. However, the mapping between energy and \textbf{k} is not always one to one.

\begin{figure}[h]
\centering
\includegraphics[trim=20 265 25 265,clip,width=1.2\linewidth]{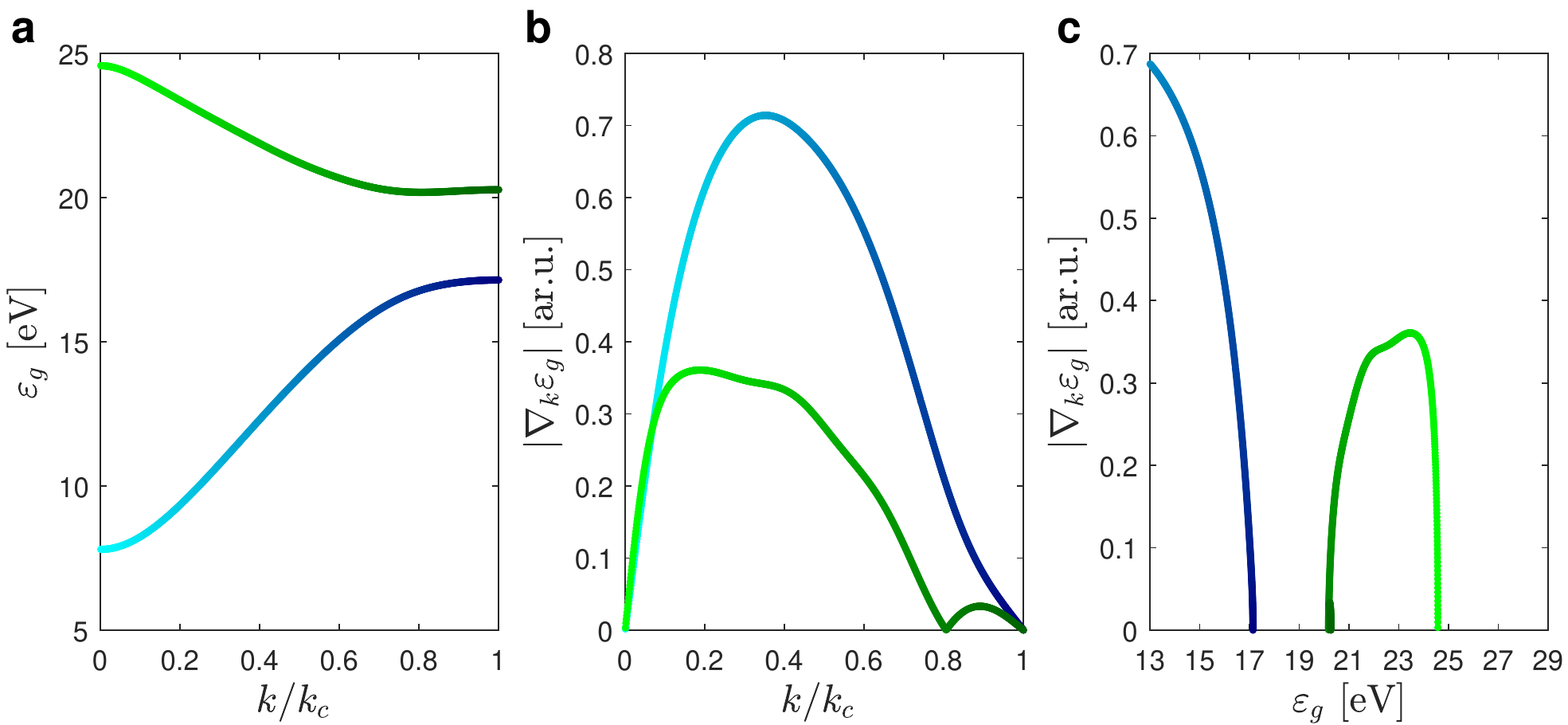}
\caption{\textbf{a} The energy gap between the valence band and the contributing conduction bands, $\varepsilon_{gn}=E_{c_n}-E_{v}$, as a function of \textbf{k}, at $0^\circ$ orientation of the crystal. The blue plot represents the gap of conduction band 5 and the green plot represents the gap of band 9 (as named in the main text, at Figure 2a). \textbf{b} The gradient of the band gap, $\nabla_{\textbf{k}}\varepsilon_{gn}$, as a function of \textbf{k}. \textbf{c} The gradient of the band gap as a function of the energy gap. In \textbf{a},\textbf{b},\textbf{c} the lighter colors in each curve stand for small values of \textbf{k} while the darker colors for large values of \textbf{k}.}
\label{fig:sifig2}
\end{figure}

In Figure \ref{fig:sifig2} we plot this mapping for $0^\circ$ orientation angle of the crystal. Figure \ref{fig:sifig2}\textbf{a} presents the energy gap between the valence band and the two contributing conduction bands (bands 5 and 9 in main text, Figure 2a) as a function of the crystal momentum (\textbf{k}). One can see that at this angle there is a one to one mapping between energy and crystal momentum, for both of the conduction bands. In Figure \ref{fig:sifig2}\textbf{b} we plot the gradient of the gap, $\nabla_{\textbf{k}}\varepsilon_g$, as a function of \textbf{k}. Using the mapping between \textbf{k} and energy we plot the gradient as a function of energy in Figure \ref{fig:sifig2}\textbf{c}.

\begin{figure}[h]
\centering
\includegraphics[trim=20 265 25 265,clip,width=1.2\linewidth]{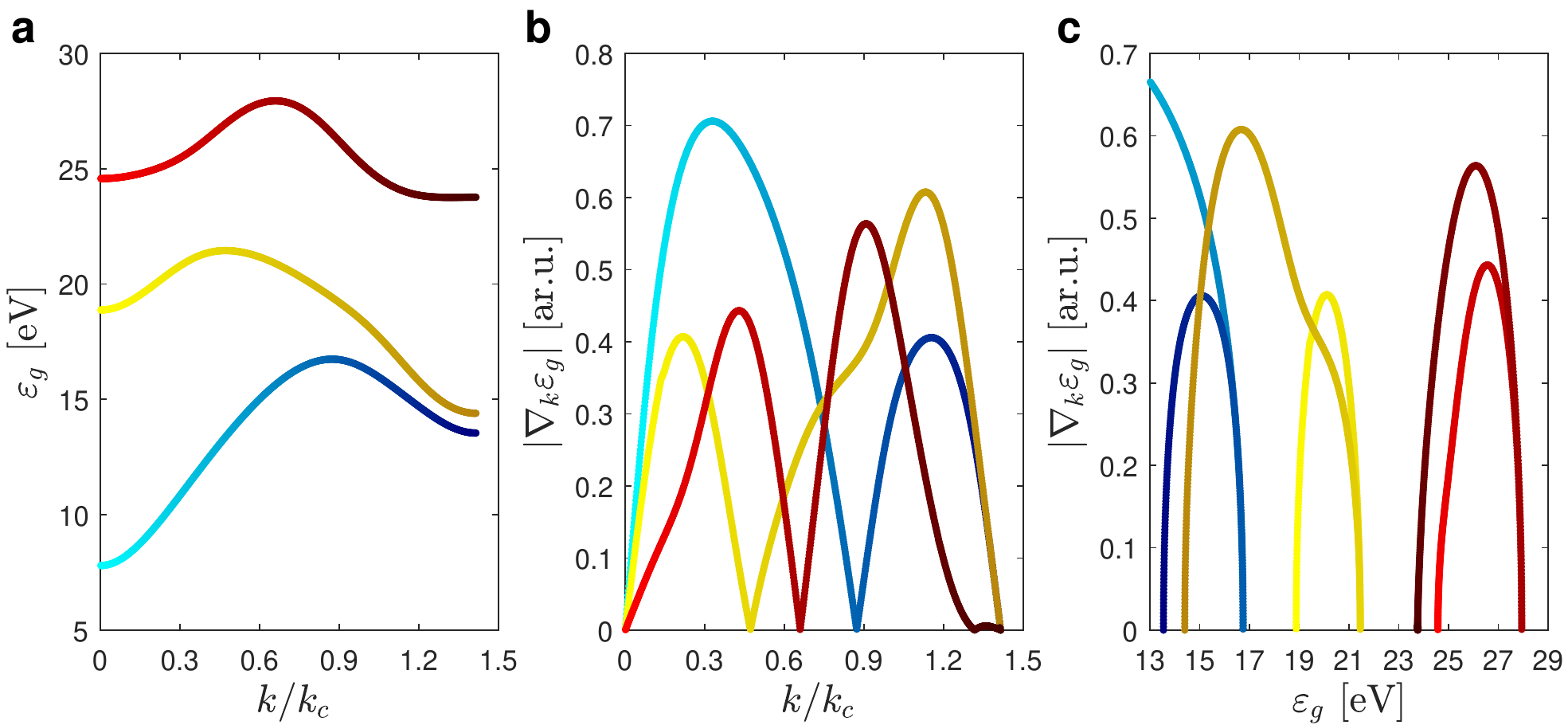}
\caption{\textbf{a} 
		The energy gap between the valence band and the contributing conduction bands, $\varepsilon_{gn}=E_{c_n}-E_{v}$, as a function of \textbf{k}, at $45^\circ$ orientation of the crystal.The blue plot represents the gap of conduction band 5, the yellow plot represents the gap of band 7 and the red plot the gap of conduction band 10 (as named in the main text, at Figure 2d). \textbf{b} The gradient of the band gap, $\nabla_{\textbf{k}}\varepsilon_{gn}$, as a function of \textbf{k}. \textbf{c} The gradient of the band gap as a function of the energy gap. In \textbf{a},\textbf{b},\textbf{c} the lighter colors in each curve stand for small values of \textbf{k} while the darker colors for large values of \textbf{k}.}
\label{fig:sifig3}
\end{figure}

In other orientations the mapping between momentum and energy is more complex. Consider, for example, the mapping for angle $45^\circ$ (Figure \ref{fig:sifig3}). At this angle, none of the contributing bands (5,7,10) has injective energy-momentum mapping as presented in Figure \ref{fig:sifig3}\textbf{a}-- there is more than one value of \textbf{k} that is attributed to the same harmonic energy. However, at the intensity range used in this experiment, we estimate that the maximum crystal momentum in which the electron could reach, $k_{max}/k_c$, is approximately 0.9-1.05 ($k_{max}=2E_{0}/\omega$). Therefore, we do not consider in figures 2b and 2e of the main text the critical points attributed to high values of \textbf{k}. 

\section{The dipole coupling}
\subsection{MgO Symmetries}
\begin{figure}[h]
 	\centering
 	\includegraphics[trim=190 35 290 30,clip,width=1\linewidth]{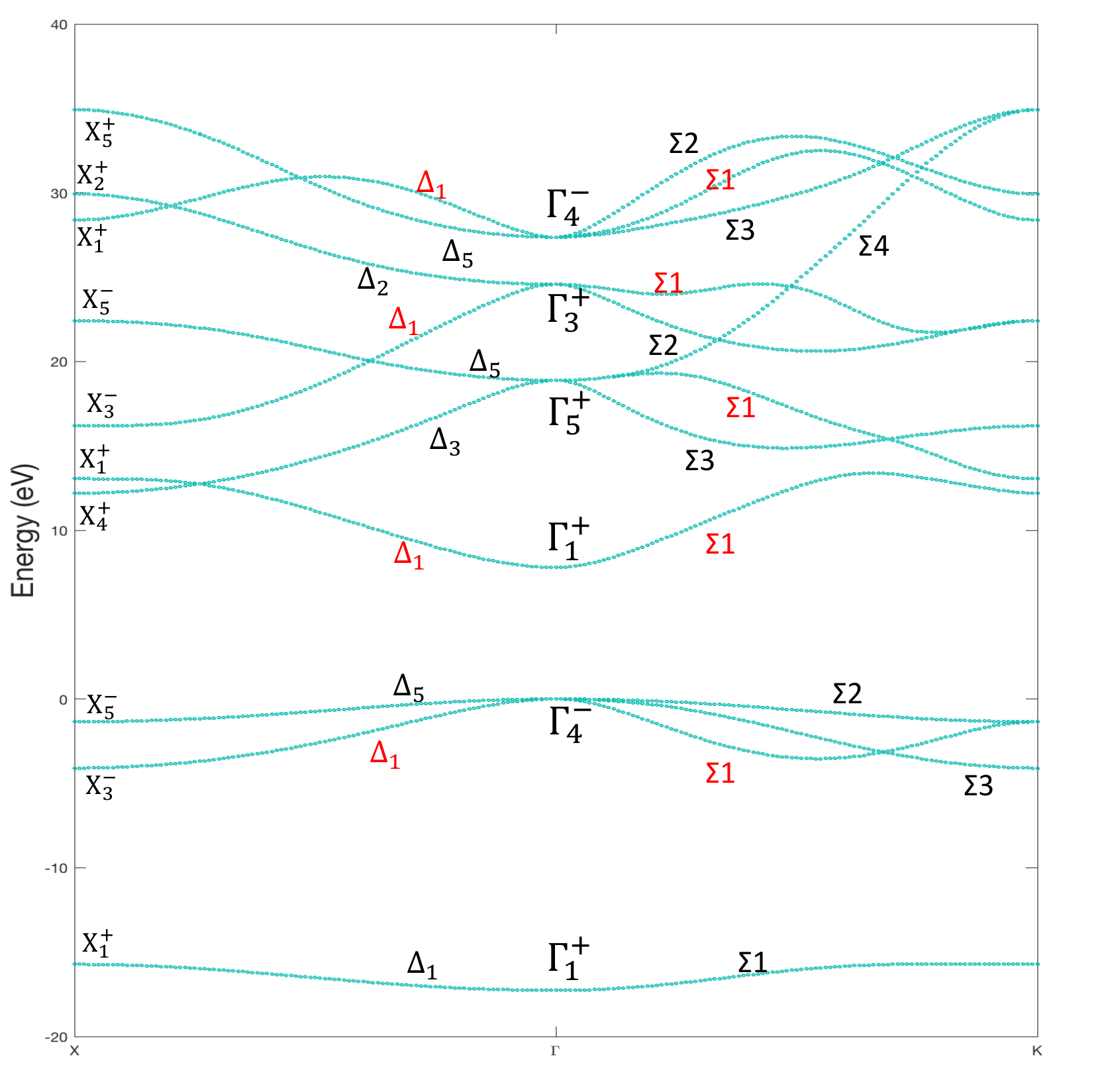}
 	\caption{The MgO band structure, along $\Gamma-X$ and $\Gamma-K$ lines, labeled by the symmetry group representations. The red labels indicate the relevant coupled bands.}
 	\label{fig:sifig4}
 \end{figure}

The lattice of MgO belongs to the symmorphic space group $Fm\overline{3}m$ (No. 225) that includes the point group $O_h$ and lattice translation. The energy bands are generally labeled by the corresponding space group representations, as shown in Figure \ref{fig:sifig4}\textbf{a}, to characterize the symmetry information of the associated electronic wavefunctions. If two bands have the same representation, their associated wavefunctions share the same eigenvalue under the symmetry operation in the space group. This is a direct outcome of the commutation relation between the Hamiltonian and the symmetry operators ($\hat{R}$) in the group: $\hat{R}^{-1}\hat{H}(\vec{r})\hat{R}=\hat{H}(\vec{r})$. However, in the crystal momentum space, a symmetry operation defined by $\hat{R}$ operator satisfies $\hat{R}^{-1}\hat{H}(\vec{k})\hat{R}=\hat{H}(\hat{R}\vec{k})$, where $\hat{H}(\vec{k})$ is the Hamiltonian in $\vec{k}$ space.
 Because $\hat{R}\vec{k}$ is not always equivalent to $\vec{k}$ (it can be equal or different by an integer number of reciprocal lattice vectors), the symmetry of the energy bands depends on $\vec{k}$ and $\hat{H}(\vec{k})$ and usually exhibits less symmetries at a generic $\vec{k}$ (called little group in the literature). At the $\Gamma$ point, for example, $\hat{H}(\vec{k})$ has all symmetries of $O_h$ since $\hat{R}\vec{k}\equiv\vec{k}$. Accordingly, the 9th and 10th bands are classified into the doubly-degenerate $\Gamma_{3}^+$ representation (see figure \ref{fig:sifig4}). However, along the $\Gamma-X$ line, this equality is not valid ($\hat{R}\vec{k}\neq\vec{k}$) and the $O_h$ group is reduced to $C_{4v}$. Due to this symmetry reduction, $\Gamma_{3}^+$ bands split into $\Delta_1$ and $\Delta_2$ along $\Gamma-X$. In a similar way, $\Gamma_{3}^+$ bands split into $\Sigma_1$ and $\Sigma_2$ along $\Gamma-K$ where the symmetry is reduced to $C_{2v}$.
 
 The bands representation is a convenient tool in the understanding of the dipole coupling between two bands. The dipole matrix element is $ \matrixelement{\Psi_{i}(\vec{k})}{\hat{x}}{\Psi_{f}(\vec{k})}$, where ${\hat{x}}$ is the position operator, $\ket{\Psi_{i}(\vec{k})}$ and $\ket{\Psi_{f}(\vec{k})}$ are initial and final states. Since the symmetry of the system depends on the crystal momentum, the selection rules for optical transitions depend as well. At the $\Gamma$ point, the system exhibits all the symmetries of the $O_h$ group including the inversion symmetry. Therefore the selection rules are similar to an ordinary optical transition in a molecular system: since the electronic wave-functions have defined parity, the initial and final states should obey opposite parity (in this point group the electric field has odd parity). One can see in figure \ref{fig:sifig4} that the allowed transitions at the $\Gamma$ point are between any bands named with plus and minus upper index ($\Gamma_{i}^{+}\rightleftarrows\Gamma_{j}^{-}$). Next, we consider the dipole coupling at a non-zero k, first in the case where the crystal is oriented at $0^\circ$ angle (k is some point along $\Gamma-X$ axis). Moving from the $\Gamma$ point, the inversion symmetry is broken and the parity of the associated wave-function is not well defined. The symmetry point is reduced to $C_{4v}$, such that the system is now defined by one rotation axis. Since the dipole operator is parallel to the rotation axis, in this symmetry group, the dipole operator is an \textit{even} function and the allowed transitions are between bands with same representation. In figure \ref{fig:sifig4} the allowed transitions along $\Gamma-X$ are between $\Delta_i$ and $\Delta_j$ where $i=j$. Since the representation of the valence band is $\Delta_1$, only bands 5 and 9 are coupled to it. In a similar way, at $45^\circ$ crystal orientation, transitions along $\Gamma-K$ axis are between bands with the same representation. In this orientation, the valence band is coupled only to bands 5, 7 and 10.

\subsection{Dipole Coupling Variation}
\begin{figure}[h]
	\centering
	\includegraphics[trim= 100 230 100 230,clip,width=1.2\linewidth]{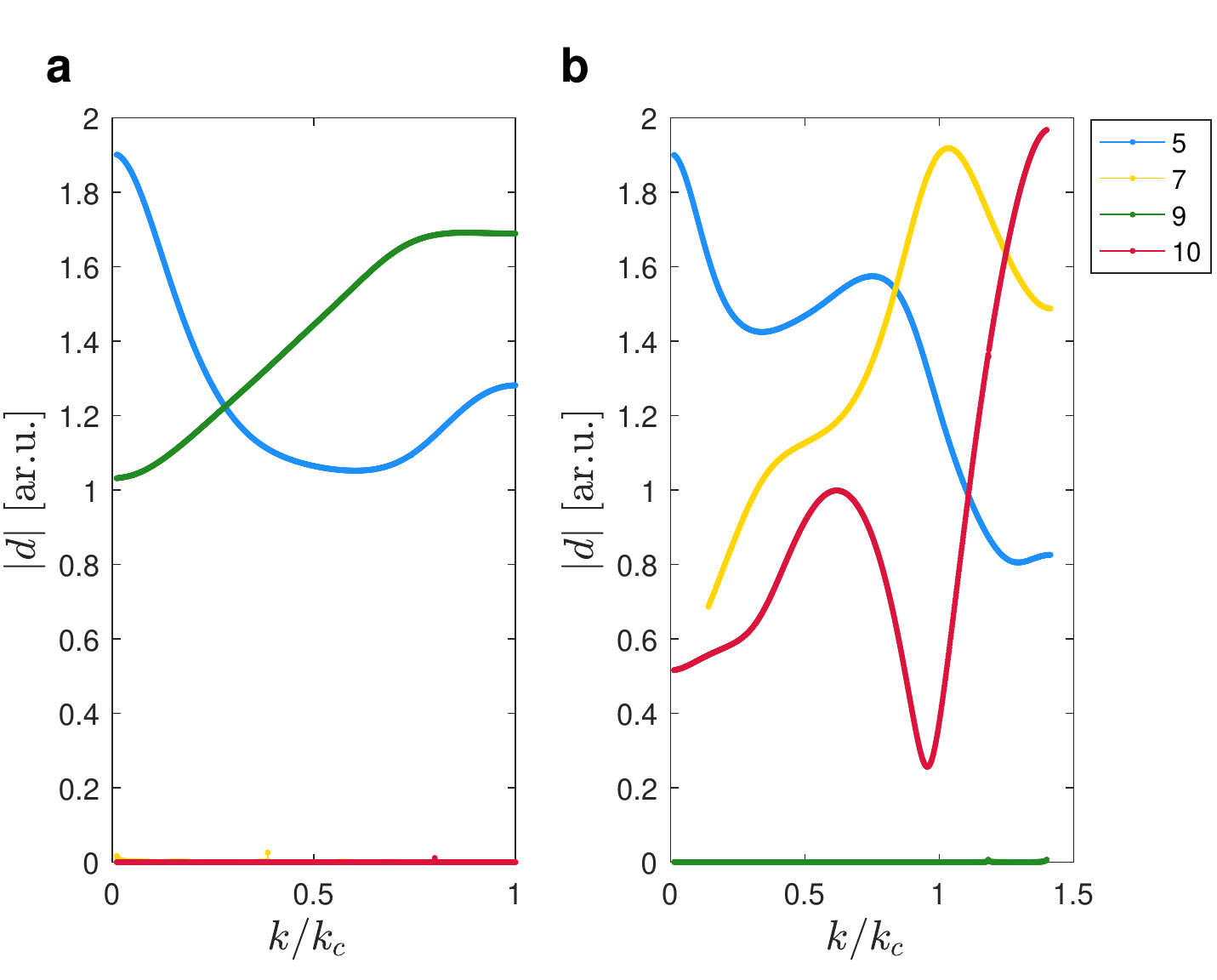}
	\caption{The absolute values of the dipole coupling between the valence band and conduction bands 5, 7, 9 and 10 as a function of the crystal momentum. \textbf{a} presents the coupling at $0^\circ$ crystal orientation. \textbf{b} presents the coupling at $45^\circ$ crystal orientation.}
	\label{fig:sifig5}
\end{figure}

 The enhanced features in the harmonic spectra and their drastic dependence on the crystal orientation cannot be explained by the energy dependence of the dipole coupling between the bands. In figure \ref{fig:sifig5} we present the dependence of the dipole couplings between the valence band and conduction bands 5, 7, 9 and 10 on the crystal momentum. Figure \ref{fig:sifig5}\textbf{a} describes the coupling for $0^\circ$ crystal orientation. At this angle only bands 5 and 9 couple to the valence band. The magnitude of their coupling changes at most by a factor of 2 through the entire range. Figure \ref{fig:sifig5}\textbf{b} presents a similar picture for the $45^\circ$ crystal orientation- the dipole couplings to the valence band span less than an order of magnitude for the entire range. Furthermore, comparing figures \ref{fig:sifig5}\textbf{a} and \ref{fig:sifig5}\textbf{b} we find that the dipoles' magnitudes are comparable. This dependence is typical for all angles and therefore cannot explain the strong angular dependence of the harmonic spectrum. Moreover, the total change in the dipole magnitude is much lower compared to the enhancement factor observed in the HHG spectrum which reaches almost 35 between the $17^{th}$ and $15^{th}$ harmonics.

 \bibliographystyle{naturemag}
 \bibliography{bib_MgO}